\documentstyle[12pt]{article}
\begin{document}
\bibliographystyle{unsrt} 
\vspace{25mm}

\begin{center}
{\large \bf TWO DIFFERENT SQUEEZE \\[2mm]
TRANSFORMATIONS}\\[7mm]
D. Han\\{\it National Aeronautics and Space Administration, Goddard Space
Flight Center,\\ Code 910.1, Greenbelt, Maryland 20771, U.S.A.}\\[5mm]
Y. S. Kim\\{\it Department of Physics, University of Maryland,
\\College Park, Maryland 20742, U.S.A.}\\[5mm]
\end{center}

\vspace{2mm}

\begin{abstract}

Lorentz boosts are squeeze transformations.  While these transformations
are similar to those in squeezed states of light, they are fundamentally
different from both physical and mathematical points of view.  The
difference is illustrated in terms of two coupled harmonic oscillators,
and in terms of the covariant harmonic oscillator formalism.

\end{abstract}

The word ``squeezed state'' is relatively new and was developed in
quantum optics, and was invented to describe a set of two photon coherent
states \cite{yuen76}.
However, the geometrical concept of squeeze or squeeze transformations has
been with us for many years.  As far as the present authors can see, the
earliest paper on squeeze transformations was published by Dirac in 1949
\cite{dir49}, in which he showed that Lorentz boosts are squeeze
transformations.  In this report, we show that Dirac's Lorentz squeeze is
different from the squeeze transformations in the squeezed state of light.
The question then is how different they are.  In order to answer this
question, we shall use a system of two coupled harmonic oscillators.

Let us look at a phase-space description of one simple harmonic oscillator.
Its orbit in phase space is an ellipse.  This ellipse can be canonically
transformed into a circle.  The ellipse can also be rotated in phase space
by canonical transformation.  This combined operation is dictated by a
three-parameter group $Sp(2)$ or the two-dimensional symplectic group.  The
group $Sp(2)$ is locally isomorphic to $SU(1,1), O(2,1)$, and $SL(2,r)$,
and is applicable to many branches of physics.  Its most recent application
was to single-mode squeezed states of light \cite{yuen76,knp91}.

Let us next consider a system of two coupled oscillators.  For this system,
our prejudice is that the system can be decoupled by a coordinate rotation.
This is not true, and the diagonalization requires a squeeze transformation
in addition to the rotation applicable to two coordinate variables
\cite{knp91,gold80}.  This is also a transformation of the symplectic
group $Sp(2)$.

If we combine the $Sp(2)$ symmetry of mode coupling and the $Sp(2)$ symmetry
in phase space, the resulting symmetry is that of the (3 + 2)-dimensional
Lorentz group \cite{dir63}.  Indeed, it has been shown that this is the
symmetry of two-mode squeezed states \cite{bishop88,hkn90}.  It is
known that the (3 + 2)-dimensional Lorentz group is locally isomorphic to
Sp(4) which is the group of linear canonical transformations in the
four-dimensional phase space for two coupled oscillators.  These canonical
transformations can be translated into unitary transformations
in quantum mechanics \cite{hkn90}.

In addition, for the two-mode problem, there is another Sp(2) transformation
resulting from the relative size of the two phase spaces.  In classical
mechanics, there are no restrictions on the area of phase space within the
elliptic orbit in phase space of a single harmonic oscillator.  In quantum
mechanics, however, the minimum phase-space size is dictated by the
uncertainty relation.  For this reason, we have to adjust the size of phase
space before making a transition to quantum mechanics.  This adds another
$Sp(2)$ symmetry to the coupled oscillator system \cite{hkn95jm}.  However,
the transformations
of this $Sp(2)$ group are not necessarily canonical, and there does not
appear to be a straightforward way to translate this symmetry group into the
present formulation of quantum mechanics.  We shall return to this problem
later in this report.

If we combine this additional $Sp(2)$ group with the above-mentioned $O(3,2)$,
the total symmetry of the two-oscillator system becomes that of the group
$O(3,3)$, which is the Lorentz group with three spatial and three time
coordinates.  This was a rather unexpected result and its mathematical
details have been published recently by the present authors \cite{hkn95jm}.
This $O(3,3)$ group has fifteen parameters and is isomorphic to $SL(4,r)$.
It has six $Sp(4)$-like subgroups and many $Sp(2)$ like subgroups.


Let us consider a system of two coupled harmonic oscillators.  The
Lagrangian for this system is
\begin{equation}\label{lagran0}
L = {1\over 2}\left\{m_{1}\dot{x}^{2}_{1} + m_{2}\dot{x}^{2}_{2} -
A' x^{2}_{1} + B' x^{2}_{2} + C' x_{1} x_{2} \right\},
\end{equation}
with
\begin{equation}
A' > 0, \qquad B' > 0, \qquad 4A'B' - C'^2 > 0 .
\end{equation}
Then the traditional wisdom from textbooks on classical mechanics is
to diagonalize the system by solving the eigenvalue equation
\begin{equation}
\left | \begin{array}{cc}A'-m_{1}\omega^2 & C' \\
C' & B' - m_{2}\omega^2 \end{array} \right | = 0 .
\end{equation}
There are two solutions for $\omega^{2}$, and these solutions indeed give
correct frequencies for the two normal modes.  Unfortunately, this
computation does not lead to a complete solution to the diagonalization
problem.  The above eigenvalue equation seems similar to that for the
rotation, but it is not.

Let us go back to Eq.(\ref{lagran0}).  This quadratic form cannot be
diagonalized by rotation alone.  Indeed, the potential energy portion of
the Lagrangian can be diagonalized by one rotation, but this rotation will
lead to a non-diagonal form for the kinetic energy.  For this reason, we
first have to replace $x_{1}$ and $x_{2}$ by $y_{1}$ and $y_{2}$ with the
transformation matrix
\begin{equation}\label{xtoy}
\pmatrix{x_{1} \cr x_{2}} = \pmatrix{(m_{2}/m_{1})^{1/4}  & 0
\cr 0 & (m_{1}/m_{2})^{1/4}} \pmatrix{y_{1} \cr y_{2}} .
\end{equation}
In terms of these new variables, the Lagrangian can be written as
\begin{equation}\label{lagran1}
L = {\sqrt{m_{1}m_{2}}\over 2}\left\{\dot{y}^{2}_{1} +
\dot{y}^{2}_{2} \right\} -
{1 \over 2}\left\{A y^{2}_{1} + B y^{2}_{2} + C y_{1} y_{2} \right\},
\end{equation}
with
$$
\pmatrix{A \cr B \cr C} = \pmatrix{\sqrt{m_{2}/m_{1}}  & 0 & 0
\cr 0 & \sqrt{m_{1}/m_{2}} & 0 \cr 0 & 0 & 1 } \pmatrix{A' \cr B' \cr C'} .
$$
The Lagrangian of Eq.(\ref{lagran1}) can now be diagonalized by a simple
coordinate rotation:
\begin{equation}\label{rotation}
\pmatrix{z_{1} \cr z_{2}} = \pmatrix{\cos\alpha & \sin\alpha
\cr -\sin\alpha & \cos\alpha} \pmatrix{y_{1} \cr y_{2}},
\end{equation}
with
\begin{equation}\label{eq.34}
\tan(2\alpha) = {C\over A - B} .
\end{equation}
In this Lagrangian formalism, momenta are not independent variables.
They are strictly proportional to their respective coordinate variables.
When the coordinates are rotated by the matrix of Eq.(\ref{rotation}),
the momentum variables are transformed according to the same matrix.  When
the coordinates undergo the scale transformation of Eq.(\ref{xtoy}), the
momentum variables are transformed by the same matrix.  Thus, the phase-space
volume is not preserved for each coordinate.

Let us approach the same problem using the Hamiltonian
\begin{equation}\label{hamil0}
H = {1\over 2}\left\{{p^{2}_{1}\over m_{1}} + {p^{2}_{2}\over m_{2}} +
A' x^{2}_{1} + B' x^{2}_{2} + C' x_{1} x_{2} \right\}.
\end{equation}
Here again, we have to rescale the coordinate variables.  In this formalism,
the central issue is the canonical transformation, and the phase-space volume
should be preserved for each mode.  If the coordinate variables are to be
transformed according to Eq.(\ref{xtoy}), the transformation matrix for the
momenta should be the inverse of the matrix given in Eq.(\ref{xtoy}).
Indeed, if we adopt this transformation matrix, the new Hamiltonian becomes
\begin{equation}\label{hamil1}
H = {1\over 2\sqrt{m_{1}m_{2}}}\left\{p^{2}_{1} + p^{2}_{2} \right\} +
{1\over 2}\left\{A x_{1}^{2} + B x^{2}_{2} + C x_{1} x_{2} \right\} .
\end{equation}
As for the rotation, the rules of canonical transformations dictate that
both the coordinate and momentum variables have the same rotation matrix.
The above Hamiltonian can be diagonalized by the rotation matrix given
in Eq.(\ref{rotation}).

We can now consider the four-dimensional phase space consisting of variables
in the following order.
\begin{equation}\label{coord1}
\left(\chi_{1}, \chi_{2}, \chi_{3}, \chi_{4} \right) =
\left(x_{1}, x_{2}, p_{1}, p_{2} \right) .
\end{equation}
For both the non-canonical Lagrangian system and the canonical Hamiltonian
system, the mode-coupling rotation matrix is
\begin{equation}\label{rot}
R(\alpha) = \pmatrix{\cos\alpha & \sin\alpha & 0 & 0 \cr
-\sin\alpha & \cos\alpha & 0 & 0 \cr
0 & 0 & \cos\alpha & \sin\alpha \cr
0 & 0 & -\sin\alpha & \cos\alpha} .
\end{equation}
On the other hand, they have different matrices for the scale transformation.
For the canonical Hamiltonian system, the matrix takes the form
\begin{equation}\label{sq-}
S_{-}(\eta) = \pmatrix{e^{\eta}  & 0 & 0 & 0 \cr
0 & e^{-\eta}  & 0 & 0 \cr
0 & 0 & e^{-\eta}  & 0 \cr
0 & 0 & 0 & e^{\eta} } .
\end{equation}
Here, the position and momentum variables undergo anti-parallel squeeze
transformations.  On the other hand, for non-canonical Lagrangian system,
the squeeze matrix is written as
\begin{equation}\label{sq+}
S_{+}(\eta) = \pmatrix{e^{\eta}  & 0 & 0 & 0 \cr
0 & e^{-\eta}  & 0 & 0 \cr
0 & 0 & e^{\eta}  & 0 \cr
0 & 0 & 0 & e^{-\eta} } .
\end{equation}
We use the notation $S_{+}$ and $S_{-}$ for the parallel and anti-parallel
squeeze transformation respectively.


If we rotate the above squeeze matrices by $45^{o}$ using the
rotation matrix of Eq.(\ref{rot}), the anti-parallel squeeze matrix
become
\begin{equation}\label{sq--}
S_{-}(\eta) = \pmatrix{\cosh\eta  & \sinh\eta & 0 & 0 \cr
\sinh\eta & \cosh\eta  & 0 & 0 \cr
0 & 0 & \cosh\eta  & -\sinh\eta \cr
0 & 0 & -\sinh\eta & \cosh\eta } ,
\end{equation}
and the parallel squeeze matrix takes the form
\begin{equation}\label{sq++}
S_{+}(\eta) = \pmatrix{\cosh\eta  & \sinh\eta & 0 & 0 \cr
\sinh\eta & \cosh\eta  & 0 & 0 \cr
0 & 0 & \cosh\eta  & \sinh\eta \cr
0 & 0 & \sinh\eta & \cosh\eta } .
\end{equation}
Now the difference between these two matrices is quite clear.
The squeeze matrix of Eq.(\ref{sq--}) is applicable to two-mode squeezed
states of light \cite{hkn90,cav85,ymk86}.

As for the squeeze matrix of Eq.(\ref{sq++}), let us consider the Lorentz
transformation of a particle along the $z$ direction:
\begin{equation}
z' = (\cosh\eta) z + (\sinh\eta) t , \qquad
t' = (\sinh\eta) z + (\cosh\eta) t .
\end{equation}
Then the momentum and energy are transformed according to
\begin{equation}
P' = (\cosh\eta) P + (\sinh\eta) E , \qquad
E' = (\sinh\eta) P + (\cosh\eta) E .
\end{equation}
If we regard $z$ and $t$ as the two coordinate variables, the
four-component vector of Eq.(\ref{coord1}) takes the form
\begin{equation}
\left(\chi_{1}, \chi_{2}, \chi_{3}, \chi_{4} \right) = (z, t, P, E ) .
\end{equation}
Thus, the parallel squeeze matrix performs a Lorentz boost.  According
to classical mechanics of coupled harmonic oscillators, this
transformation appears like a non-canonical transformation.  Then, is
the Lorentz boost a non-canonical transformation?  The answer is NO.


We would like to show that the Lorentz boost is an uncertainty-preserving
transformation using the covariant oscillator formalism which has been
shown to be effective in explaining the basic hadronic features observed
in high energy laboratories \cite{knp86}.  According to this model, the
ground-state wave function for the hadron takes the form
\begin{equation}\label{wf1}
\psi_{0}(z,t) = \left({1 \over \pi }\right)^{1/2}
\exp\left\{-{1\over 2}\left(z^{2} + t^{2}\right)\right\} ,
\end{equation}
where the hadron is assumed to be a bound state of two quarks, and $z$
and $t$ are space and time separations between the quarks.  If the system
is boosted, the wave function becomes \cite{knp86}
\begin{equation}\label{wf2}
\psi _{\eta }(z,t) = \left({1 \over \pi }\right)^{1/2} 
\exp \left\{-{1\over 2}\left(e^{-2\eta }u^{2} + 
e^{2\eta }v^{2}\right)\right\} ,
\end{equation}
where
$$
u = (z + t)/\sqrt{2}, \qquad v = (z - t)/\sqrt{2} .
$$
The $u$ and $v$ variables are called the light-cone variables \cite{dir49}.
The wave function of Eq.(\ref{wf1}) is distributed
within a circular region in the $u v$ plane, and thus in the $z t$ plane.  
On the other hand, the wave function of Eq.(\ref{wf2}) is distributed in
an elliptic region.  This ellipse is a ``squeezed'' circle with the same
area as the circle.  The question then is how the momentum-energy
wave function is squeezed.

The momentum wave function is obtained from the Fourier transformation of
the expression given in Eq.(\ref{wf2}):
\begin{equation}\label{fourier}
\phi_{\eta }(q_{z},q_{0}) = \left({1 \over 2\pi }\right) 
\int \psi_{\eta}(z, t) \exp{\left\{-i(q_{z}z - q_{0}t)\right\}} dx dt .
\end{equation}
If we use the variables:
\begin{equation}\label{conju}
q_{u} = (q_{0} - q_{z})/\sqrt{2} , \qquad q_{v} = (q_{0} + q_{z})/\sqrt{2} .
\end{equation}
In terms of these variables, the above Fourier transform can be
written as
\begin{equation}\label{wf3}
\phi_{\eta }(q_{z},q_{0}) = \left({1 \over 2\pi }\right) 
\int \psi_{\eta}(z, t) \exp{\left\{-i(q_{u} u + q_{v} v)\right\}} du dv .
\end{equation}
The resulting momentum-energy wave function is 
\begin{equation}\label{phi}
\phi_{\eta }(q_{z},q_{0}) = \left({1 \over \pi }\right)^{1/2} 
\exp\left\{-{1\over 2}\left(e^{-2\eta}q_{u}^{2} + 
e^{2\eta}q_{v}^{2}\right)\right\} .
\end{equation}
Because we are using here the harmonic oscillator, the mathematical form 
of the above momentum-energy wave function is identical with that of the
space-time wave function given in Eq.(\ref{wf2}).  The Lorentz-squeeze
properties of these wave functions are also the same.  This certainly
is consistent with the parallel squeeze matrix given in Eq.(\ref{sq++}),
and the Lorentz boosts appears like a non-canonical transformation.

However, we still have to examine how conjugate pairs are chosen from
the space-time and momentum-energy wave functions.  Let us go back to
Eq.(\ref{fourier}) and
Eq.(\ref{wf3}).  It is quite clear that the light-cone variable $u$ and
$v$ are conjugate to $q_{u}$ and $q_{v}$ respectively.  It is also clear 
that the distribution along the $q_{u}$ axis shrinks as the $u$-axis 
distribution expands.  The exact calculation leads to 
\begin{equation}
<u^{2}><q_{u}^{2}> = 1/4 , \qquad  <v^{2}><q_{v}^{2}> = 1/4  .    
\end{equation}
Planck's constant is indeed a Lorentz-invariant quantity, and the Lorentz
boost is a canonical transformation.

Because of the Minkowskian metric we used in the Fourier transformation
of Eq.({\ref{fourier}), the non-canonical squeeze transformation of
Eq.(\ref{sq++}) becomes a canonical transformation for the Lorentz
boost.  Otherwise, it remains non-canonical.  Then, does this
non-canonical transformation play a role in physics?  The answer is
YES.  The best known examples are thermally excited oscillator
states \cite{ume82} and coupled oscillator system where one of the
oscillator is not observed \cite{yupo87,ekn89}.  These systems serve
as simple models for studying the role of entropy in quantum
mechanics \cite{neu32,fey72}.

These examples are for the cases where the phase space volume for
each mode becomes larger than Planck's constant.  In the classical
mechanics of two coupled harmonic oscillators, the phase-space volume
of each oscillator fluctuates.  If one becomes larger, the other
shrinks.  In quantum mechanics, we do not have a theory of shrinking
phase-space volumes.  Without this, we cannot have a complete
understanding of coupled oscillators in quantum mechanics.

\end{document}